\documentclass[10pt,prl,twocolumn,superscriptaddress]{revtex4}

\usepackage{graphicx}
\usepackage{amssymb}
\usepackage{times}
\usepackage{amsmath}
\usepackage{amsfonts}
\usepackage{txfonts}
\usepackage{color}

\begin{document}

\title{Cavity Optomechanical Magnetometer}

\author{S. Forstner}  \affiliation{School of Mathematics and Physics, University of Queensland, St Lucia, Queensland 4072, Australia}
\affiliation{Physik-Department, TU Muenchen, 85748 Garching, Germany}

\author{S. Prams} \affiliation{School of Mathematics and Physics, University of Queensland, St Lucia, Queensland 4072, Australia}

\author{J. Knittel} \affiliation{School of Mathematics and Physics, University of Queensland, St Lucia, Queensland 4072, Australia}\affiliation{Centre for Engineered Quantum Systems, University of Queensland, St Lucia, Brisbane, QLD 4072, Australia}

\author{E. D. van Ooijen} \affiliation{School of Mathematics and Physics, University of Queensland, St Lucia, Queensland 4072, Australia}

\author{J. D. Swaim} \affiliation{School of Mathematics and Physics, University of Queensland, St Lucia, Queensland 4072, Australia} \affiliation{Centre for Engineered Quantum Systems, University of Queensland, St Lucia, Brisbane, QLD 4072, Australia}

\author{G. I. Harris} \affiliation{School of Mathematics and Physics, University of Queensland, St Lucia, Queensland 4072, Australia} \affiliation{Centre for Engineered Quantum Systems, University of Queensland, St Lucia, Brisbane, QLD 4072, Australia}

\author{A. Szorkovszky} \affiliation{School of Mathematics and Physics, University of Queensland, St Lucia, Queensland 4072, Australia} \affiliation{Centre for Engineered Quantum Systems, University of Queensland, St Lucia, Brisbane, QLD 4072, Australia}

\author{W. P. Bowen}\affiliation{School of Mathematics and Physics, University of Queensland, St Lucia, Queensland 4072, Australia} \affiliation{Centre for Engineered Quantum Systems, University of Queensland, St Lucia, Brisbane, QLD 4072, Australia}

\author{H. Rubinsztein-Dunlop} \affiliation{School of Mathematics and Physics, University of Queensland, St Lucia, Queensland 4072, Australia} \affiliation{Centre for Engineered Quantum Systems, University of Queensland, St Lucia, Brisbane, QLD 4072, Australia}

\begin{abstract}
A cavity optomechanical magnetometer is demonstrated. The magnetic field induced expansion of a magnetostrictive material is resonantly transduced onto the physical structure of a highly compliant optical microresonator, and read-out optically with ultra-high sensitivity. A peak magnetic field sensitivity of 400~nT~Hz$^{-1/2}$ is achieved, with theoretical modeling predicting the possibility of sensitivities below 1~pT~Hz$^{-1/2}$. This chip-based magnetometer combines high-sensitivity and large dynamic range with small size and room temperature operation.
\end{abstract}

\date{\today} \maketitle

Ultra-low field magnetometers are essential components for a wide range of practical applications including geology, mineral exploration, archaeology, defence and medicine\cite{ref1}. The field is dominated by superconducting quantum interference devices (SQUIDs) operating at cryogenic temperatures\cite{ref11}. Magnetometers capable of room temperature operation offer significant advantages both in terms of operational costs and range of applications. The state-of-the-art are magnetostrictive magnetometers with sensitivities in the range of fT~Hz$^{-1/2}$\cite{ref6, ref6b}, and atomic magnetometers which achieve impressive sensitivities as low as 160 aT~Hz$^{-1/2}$\cite{ref8} but with limited dynamic range due to the nonlinear Zeeman effect\cite{ref11, ref13}. 
%
Recently, significant effort has been made to miniaturize room temperature magnetometers. However both atomic and magnetostrictive magnetometers remain generally limited to millimeter or centimeter size scales. Smaller microscale magnetometers have many potential applications in biology, medicine, and condensed matter physics\cite{njpBouchardHall, ref19}. A particularly important application is magnetic resonance imaging, where by placing the magnetometer in close proximity to the sample both sensitivity and resolution may be enhanced\cite{Stefan_ref22_29}, potentially enabling detection of nuclear spin noise\cite{Meriles}, imaging of neural networks\cite{ref19}, and advances in areas of medicine such as magneto-cardiography\cite{ref1,ref13} and magneto-encephalography\cite{Stefan_ref25}.


In the past few years, rapid progress has been achieved on NV center based magnetometers. They combine sensitivities as low as 4 nT~Hz$^{-1/2}$ with room temperature operation, optical readout and nanoscale size\cite{ref16} and are predicted theoretically to reach the fT~Hz$^{-1/2}$ range\cite{ref17}.  This has allowed three-dimensional magnetic field imaging at the micro scale using ensembles of NV-centers \cite{ref19}, and magnetic resonance \cite{ref18} and field imaging\cite{ref17} at the nanoscale using single NV centers. 
In spite of these extraordinary achievements applications are hampered by fabrication issues and the intricacy of the read-out schemes\cite{ref20}. Furthermore miniaturization is limitied by the bulky read-out optics, the magnetic field coils for state preparation and the microwave excitation device\cite{ref19}.

In this letter we present the concept of a cavity optomechanical field sensor which combines room temperature operation and high sensitivity with large dynamic range and small size. The sensor leverages results from the emergent field of cavity optomechanics where ultra-sensitive force and position sensing has been demonstrated\cite{ref21}. A cavity optomechanical system (COMS) is used to sense magnetic field induced deformations of a magnetostrictive material, which are detected with an all-in-fiber optical system suitable for the telecom wavelength range. The presence of mechanical and optical resonances greatly enhances both the response to the magnetic field and the measurement sensitivity. 

Three implementations using different types of COMS are investigated, microscale Fabry-Perot resonators\cite{ref38}, optomechanical zipper cavities\cite{ref39}, and toroidal whispering gallery mode (TWGM) resonators\cite{ref22}, all of which are approximately 50~$\mu$m in size. Theoretical modelling predicts ultimate Brownian noise limited sensitivities at the level of 200~pT~Hz$^{-1/2}$, 10~pT~Hz$^{-1/2}$, and 700~fT~Hz$^{-1/2}$, respectively for each architecture. We experimentally demonstrate the concept with a TWGM resonator fabricated on a silicon wafer, achieving a peak sensitivity of 400~nT~Hz$^{-1/2}$. The possibility to integrate TWGM resonators in a dense two-dimensional array with waveguides for optical coupling\cite{ref24} provides the potential for highly-integrated imaging magnetometers\cite{ref23}. 

A COMS based field sensor consists of a field-sensitive material, i.e. a magnetostrictive material coupled to the COMS’s mechanical oscillator. On application of a modulated field the field-sensitive material generates an oscillating mechanical stress field within the COMS structure exciting its mechanical eigenmodes. The vibrations modulate the length of the COMS’s optical resonator. By coupling light from a laser to the cavity at or close to the wavelength of a suitable optical resonance, the vibrations are imprinted on the transmission spectrum and thus can be detected all-optically. To estimate the field sensitivity the COMS based field sensor is modelled by a forced harmonic oscillator with a single mechanical eigenmode with eigenfrequency  $\omega_m$, effective mass  $m$, and quality factor $Q$. In this case the displacement of each volume element of the COMS $\vec{u}(\vec{r},t) = x(t) \vec{u}(\vec{r})$  is given by the product of the scalar displacement  amplitude  $x(t)$, and the spatial mode shape function $\vec{u}(\vec{r})$\cite{ref28}.
\begin{figure}[t!]
\begin{center}
\includegraphics[width=8cm]{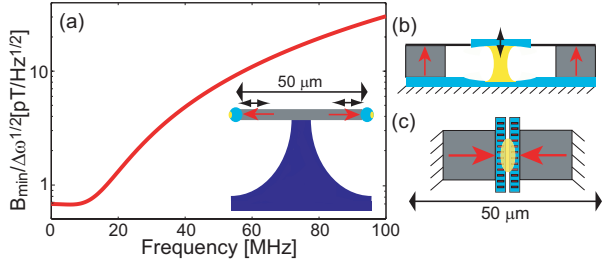}
\caption{(color online).  Maximum B-field sensitivity of three COMS based magnetometers.  a) Cross-section of a TWGM resonator consisting of a Terfenol-D disc surrounded by an optical resonator (light blue). b) Cross-section of a Fabry-Perot resonator with Terfenol-D actuators that excite vibrational modes in a suspended micro-mirror\cite{ref38}. c) Top-view of an optomechanical zipper cavity attached to two Terfenol-D actuators\cite{ref39}.  Red arrows represent movement of the Terfenol-D (in gray), black arrows mechanical vibrations.}
\label{fig1}
\end{center}
\end{figure}

In the frequency domain the motion of a forced harmonic oscillator driven at frequency  $\omega$  is given by $x(\omega) = \chi(\omega)  [F_{\rm sig}(\omega) +  F_{\rm therm}(\omega) ]$,
%
%
where $\chi(\omega)=[m (\omega_m^2 - \omega^2 + i \omega_m \omega/Q)]^{-1}$ is the mechanical susceptibility,  $F_{\rm therm}(\omega) $ the Brownian noise force with the spectral density $\langle |F_{\rm therm}(\omega) |^2\rangle = 2 k T m \omega_m/Q$\cite{ref29}, and  $F_{\rm sig}(\omega)= F_{\rm sig} \delta(\omega-\omega_{\rm sig})$ an effective harmonic driving force at the signal frequency $\omega_{\rm sig}$. This force is generated by the time-dependent mechanical stress tensor ${\bf T}(\vec{r}) e^{i \omega_{\rm sig}t}$ induced by the external magnetic field, which produces the body force density $\vec{f}_{\rm sig}(\vec{r})=\nabla \cdot {\bf T}(\vec{r})$\cite{ref30}. The strength of the effective scalar driving force $F_{\rm sig}$  experienced by the mechanical mode depends on the overlap with the spatial mode shape function\cite{ref30, ref31} and is given by
\begin{equation}
F_{\rm sig} = \int \vec{f}_{\rm sig}(\vec{r}) \cdot \vec{u}(\vec{r})  dV. \label{2}
\end{equation}
The displacement amplitude  $x(\omega)$ shifts the resonance frequency of the optical resonator $\Omega_0$ by $\delta \Omega = g x (\omega)$, where $g$ is the opto-mechanical coupling constant\cite{ref28}. In an experiment this frequency shift is imprinted on the transmission spectrum of the optical cavity and measured with a spectrum analyzer. Including the spectral density of measurement noise $S_{\delta \Omega^2}^{\rm meas}(\omega)$ due to electronic noise, laser noise and thermorefractive noise, the resulting resonance frequency shift spectrum is $S_{\delta \Omega^2} (\omega)=g^2  \langle  | x (\omega)  |^2  \rangle  + S_{\delta \Omega^2}^{\rm meas}(\omega)$. Expanding this expression we find
\begin{eqnarray}
S_{\delta \Omega^2} (\omega) \! \!
 &=& \! \! g^2 \! \left | \chi (\omega)\right |^2 \left [  F_{\rm sig}^2 \delta(\omega-\omega_{\rm sig}) \! + \! \frac{2kT m \omega_m}{Q} \right ] \! + \! S_{\delta \Omega^2}^{\rm meas} (\omega) \nonumber,
\end{eqnarray}
where the signal field is assumed to be single frequency at frequency $\omega_{\rm sig}$.
%
%
%
%
The minimum detectable force  $F_{\rm sig}^{\rm min}$ is obtained by integrating the signal and noise contributions over the bandwidth of the measuring system $\Delta \omega$  and determining the force at which the signal-to-noise ratio is unity with the result
\begin{equation}
\frac{F_{\rm sig}^{\rm min}(\omega)}{\Delta \omega^{1/2}} = \left [\frac{2 k T m \omega_m}{Q} + \frac{S_{\delta \Omega^2}^{\rm meas} (\omega) }{g^2 \left | \chi(\omega)\right |^2} \right ]^{1/2}. \nonumber 
\end{equation}
Assuming a homogeneous magnetic field $B_x$ oriented in the $x$-direction and a suitably oriented magnetostrictive medium with only a single non-zero magnetostrictive coefficient $\alpha_{\rm mag}$ which causes the material to stretch in the same direction, the magnetostrictive induced stress tensor $\bf T$  has only a single component given by $T_{xx} = \alpha_{\rm mag} B_x$.
%
Eq.~(\ref{2}) then yields the effective driving force that excites the mechanical eigenmode 
\begin{equation}
F_{\rm sig} = B_x \int \frac{\partial \alpha_{\rm mag}}{\partial x} u_x(\vec{r}) dV = B_x c_{\rm act}, \label{8}
\end{equation}
where  $u_x(\vec{r})$ is the x-component of $\vec{u}(\vec{r})$  and $c_{\rm act}$  characterizes how well the magnetic field is converted into an applied force on the oscillator and is referred to here as the magnetic actuation constant. The minimum detectable magnetic field can then finally be expressed as
\begin{equation}
\frac{B_{x}^{\rm min}(\omega)}{\Delta \omega^{1/2}} = \frac{1}{c_{\rm act}} \left [\frac{2 k T \omega_m m}{Q} + \frac{S_{\delta \Omega^2}^{\rm meas} (\omega) }{g^2 \left | \chi(\omega)\right |^2} \right ]^{1/2}. \label{9}
\end{equation}
The largest reduction is achieved at the resonance frequency $\omega_m$,  where $\left | \chi(\omega)\right |$ is maximized.
%
%
As expected, the isolation from the surrounding heat bath and the enhanced mechanical motion achieved by coupling the magnetostrictive material to an oscillator enhances the sensitivity. In the thermal noise dominated limit, the sensitivity scales as  $m^{1/2}/c_{\rm act} Q^{1/2}$, while in the measurement noise limited regime it scales as $m/c_{\rm act} Q g$.



To estimate the achievable field sensitivity we consider three realistic scenarios based on different COMS, a TWGM resonator, an optomechanical zipper cavity\cite{ref39}, and a miniature Fabry-Perot resonator with suspended micro-mirror\cite{ref38}, as depicted in Fig.~\ref{fig1}a-c. In each case, the magnetostriction is provided by the rare earth alloy Terfenol-D which 
has a relatively large magnetostrictive coefficient of $\alpha_{\rm mag} = 5 \times 10^8$~N~T$^{-1}$m$^{-2}$ at room temperature\cite{ref32, ref33}.The TWGM resonator based scenario is of most relevance to our experiments and consists of a central Terfenol-D cylinder of $24 \mu$m radius and $5 \mu$m height surrounded by a cylindrically shaped silica ring with $R = 30~\mu$m outer radius which constitutes the optical resonator. We consider a radial breathing mode with typical optomechanical coupling constant and mechanical quality factor of $g = \Omega_0/R = 6\times10^{19}$~m$^{-1}$s$^{-1}$\cite{ref28} and $Q=1000$ respectively,  where the displacement amplitude $x$ is defined to correspond to the radial displacement at the outer surface of the ring.  A finite element simulation (COMSOL) yielded the eigenfrequency $\omega_m = 2 \pi \times 5.6$~MHz  and effective mass  $m = 35$~ng\cite{ref28}, and the magnetic actuation constant $c_{\rm act} = 0.19$~N/T was obtained by solving Eq.~(\ref{8})\cite{ref35}. A measurement noise level of $S_{\delta \Omega^2}^{\rm meas}(\omega)^{1/2} = 20$~Hz/Hz$^{1/2}$\cite{ref21} was chosen at an optical resonance frequency of $\Omega_0 = 1.8 \times10^{15}$~s$^{-1}$, as is typical for TWGM resonators. Fig.~\ref{fig1}a shows the resulting magnetic field sensitivity as predicted by Eq.~(\ref{9}), with a maximum predicted sensitivity of 700~fT~Hz$^{-1/2}$  at resonance. By comparison, the peak sensitivities predicted for miniature Fabry-Perot resonators and optomechanical zipper cavities of similar spatial dimensions were 200 pT~Hz$^{-1/2}$ and 10~pT~Hz$^{-1/2}$, respectively (see Supplemental Information for calculations\cite{ref35}). 
The Fabry-Perot resonator based system was least sensitive, as the magnetostrictive material excites the mechanical modes of the suspended mirror indirectly and inefficiently by vibrating the entire mirror mount\cite{ref35}. The TWGM magnetometer shows the best sensitivity as the magnetostrictive material couples to the vibrating mode over its entire radius which results in the largest actuation constant $c_{\rm act}$  of the three systems.

\begin{figure}[t!]
\begin{center}
\includegraphics[width=8cm]{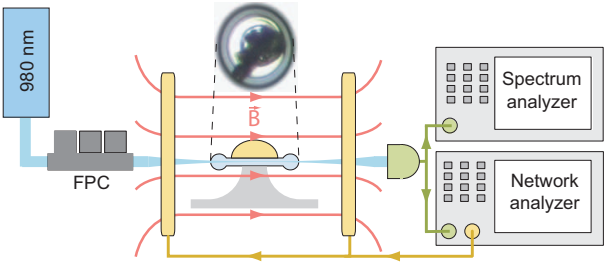}
\caption{(color online). Experimental set-up showing a TWGM resonator with a Terfenol-D sample attached.}
\label{fig2}
\end{center}
\end{figure}

As a first proof of principal demonstration of a cavity optomechanical magnetometer, a piece of Terfenol-D (Etrema Products Inc.) with a size of roughly $50\times15\times10~\mu$m$^3$ was affixed to the top surface of a TWGM resonator using micromanipulators and two-component epoxy. The TWGM resonator had major and minor diameters of 60 and 6~$\mu$m, respectively, and a 10~$\mu$m undercut\cite{ref22}. Fig.~\ref{fig2} shows the experimental set-up. The TWGM resonator was placed between two 20 mm diameter coils that generated an HF magnetic field previously calibrated with a commercial Hall sensor. 980~nm light from a widely tunable, external cavity diode laser was passed through a fibre polarization controller and evanescently coupled to the resonator via a tapered optical fibre\cite{ref29}. The laser was thermally locked to the full width half maximum of an optical resonance\cite{Terry}. Mechanical vibrations of the TWGM resonator shift its optical resonances and thus modulate the transmitted light that is detected with an InGaAs photodiode. 

\begin{figure}[t!]
\begin{center}
\includegraphics[width=7cm]{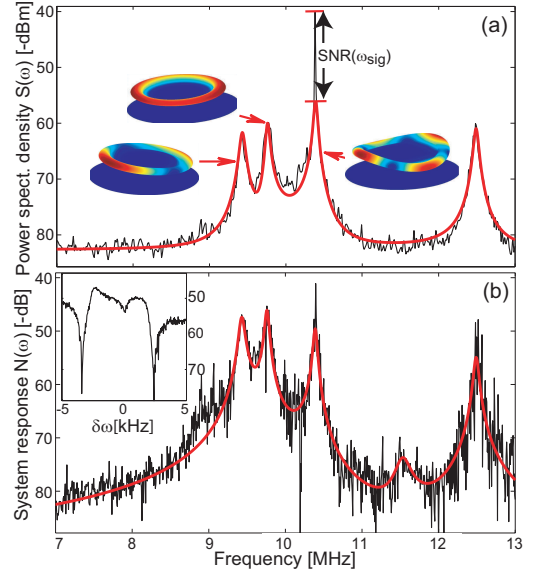}
\caption{(color online). (a) Brownian noise spectrum with magnetic excitation at 10.38 MHz. (b) System response as a function of applied B-field frequency. Inset: magnified system response centered around 10.385 MHz. Red curves: Lorentzian fits.}
\label{fig3}
\end{center}
\end{figure}

The transduction spectrum shown in Fig.~\ref{fig3}a was obtained by analyzing the transmitted light with a spectrum analyser. It shows multiple characteristic peaks corresponding to thermally excited mechanical modes. With a finite element solver the three vibrational modes around 10~MHz were identified as the lowest order flexural mode (centre) and the two lowest order crown modes. Application of a  250~$\mu$T magnetic field at  $\omega_{\rm ref} = 2 \pi \times 10.38$~MHz resulted in a signal at that frequency above the Brownian noise confirming the ability of the sensor to detect magnetic fields. The magnetic field sensitivity was determined by
\begin{equation}
B_x^{\rm min} (\omega_{\rm ref})/\Delta\omega^{1/2} = {\left | B_x \right|}/\! \!{\sqrt{{\rm SNR}(\omega_{\rm ref}) \, \Delta\omega_{\rm RBW}}}, \label{11}
\end{equation}
where SNR is the signal to noise ratio,  $B_x$ the applied magnetic field in direction $x$, and $\Delta \omega_{\rm RBW}$  the resolution bandwidth of the spectrum analyzer. 
%

To determine the sensitivity over a wide frequency range, a network analyzer was used to scan the frequency of the magnetic field from 7~to~13 MHz and simultaneously detect the system’s response $N(\omega)$.
As expected, Fig.~\ref{fig3}b shows a Lorentzian line shaped peak distribution similar to the Brownian noise spectrum (Fig.~\ref{fig3}a), with an additional peak present around 11.5~MHz, due to the enhanced noise rejection of the phase sensitive network analyzer. In both cases the measured data agrees quite well with a model with five mechanical resonances\cite{ref29}. 


Importantly, the apparent large noise amplitudes visible in Fig.~\ref{fig3}b are stationary in time, and in fact corresponds to an ensemble of resonant features having $Q$-factors on the order of $10^4$. Two such features are shown in the inset. We believe that these resonances are due to ultrasonic waves in the Terfenol-D grains that are excited by a magnetostrictive mechanism. Similar resonances have been reported for other magnetostrictive materials with comparable $Q$-factors\cite{ref41, ref42}, though to our knowledge, this is the first such observation in Terfenol-D. Constructive and destructive interferences between neighboring resonances cause strong positive and negative variations in the spectrum. It is expected that the density of resonances could be greatly reduced by engineering the geometry of the Terfenol-D. For example, the simple geometry used in Ref.~\cite{ref41} meant that only roughly two resonances per MHz were observed in agreement with theoretical predictions.

\begin{figure}[t!]
\begin{center}
\includegraphics[width=8cm]{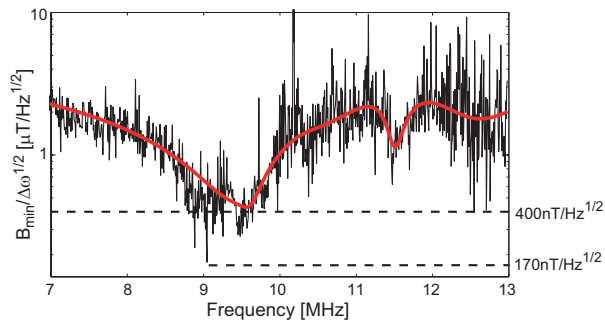}
\caption{(color online). B-field sensitivity as a function of frequency.}
\label{fig5}
\end{center}
\end{figure}

The sensitivity spectrum $B_{\rm sig}^{\rm min}(\omega)$ shown in Fig.~\ref{fig5} was obtained using the relation $B_{\rm sig}^{\rm min}(\omega) = \sqrt{S(\omega) N(\omega_{\rm ref})/N(\omega) S(\omega_{\rm ref})} B_{\rm sig}^{\rm min} (\omega_{\rm ref})$,
%
%
with the reference sensitivity obtained from a measurement similar to that in Fig.~\ref{fig3}a using Eq.~(\ref{11}). Due to the Terfenol-D resonances, the sensitivity varies between $2 \mu$T~Hz$^{-1/2}$ and 170 nT~Hz$^{-1/2}$, depending on whether destructive or constructive interference occurs. At resonance we achieve a peak sensitivity of 400~nT~Hz$^{-1/2}$ . As expected, this value is substantially above the  700~fT~Hz$^{-1/2}$ predicted for an ideal system. In our first prototype the response of the Terfenol-D was transduced to the resonator through an epoxy layer and its geometry was not well controlled. As a result the overlap between the applied force and the material motion is expected to be quite poor. Furthermore, the mechanical modes best coupled were crown modes, which themselves do not couple with high efficiency to the optical field\cite{ref28}. These limitations could be overcome by controlled deposition of Terfenol-D via sputter coating. Further sensitivity enhancement may be achieved by using TWGM resonators with higher compliance such as is achieved with spoked structures\cite{ref34}.

In conclusion, a cavity optomechanical magnetometer is reported where the magnetic field induced strain of a magnetostrictive material is coupled to the mechanical vibrations of an optomechanical resonator. A peak sensitivity of 400~nT~Hz$^{-1/2}$ is achieved, with theoretical calculations predicting sensitivities in the high fT~Hz$^{-1/2}$ 
with an optimized apparatus. This optomechanical magnetometer combines high-sensitivity, large dynamic range with small size and room temperature operation, providing an alternative to NV center based magnetometers for ultra-sensitive room temperature applications, low-field NMR and magnetic resonance imaging, and magnetic field mapping with high-density magnetometer arrays.



\emph{Acknowledgements} This research was funded by the Australian Research Council Centre of Excellence CE110001013 and Discovery Project DP0987146. Device fabrication was undertaken within the Queensland Node of the Australian Nanofabrication Facility.

\end{document}